\documentclass{webofc}\usepackage[varg]{txfonts}
\usepackage{epsfig,graphics,amssymb,amsmath,mathrsfs,color,colortbl,
bm,booktabs,cool,dcolumn}

\begin{document}\title{Semirelativistic Potential Modelling of
Bound States:\\Advocating Due Rigour}\author{Wolfgang Lucha
\inst{1}\fnsep\thanks{\email{Wolfgang.Lucha@oeaw.ac.at}}}
\institute{Institute for High Energy Physics, Austrian Academy of
Sciences, Nikolsdorfergasse 18,\\A-1050 Vienna, Austria}

\abstract{The Poincar\'e-covariant quantum-field-theoretic
description of bound states by the homogeneous Bethe--Salpeter
equation usually exhibits an intrinsic complexity that can be
attenuated by allowing this formalism to undergo various
simplifications. The resulting approximate outcome's reliability
can be assessed by applying several rigorous constraints on the
nature of the bound-state spectra; most prominent here are
existence, number and location of discrete eigenvalues.}\maketitle

\section{Our Battlefield: Semirelativistic Bound States of Least
Complexity}A popular (possibly even analytic) approach to
semirelativistic bound states of scalar particles is known under
the name spinless Salpeter equation, which is just the eigenvalue
equation of a Hamiltonian $H$ encompassing both the relativistic
kinetic term of the bound-state constituents and a static
potential $V$ encoding all interactions. For the case of two
bound-state constituents, of masses $m_1$ and $m_2$, and relative
coordinates and momenta $\bm{x}$ and $\bm{p}$, this Hamiltonian
$H$~reads\begin{equation}H=\sqrt{\bm{p}^2+m_1^2}+
\sqrt{\bm{p}^2+m_2^2}+V(\bm{x})\ ,\label{H}\end{equation}which,
for equal masses $m_1=m_2=m$ of its bound-state constituents,
simplifies (a little~bit)~to$$\widetilde H=2\,\sqrt{\bm{p}^2+m^2}+
V(\bm{x})\ .$$The spinless Salpeter equation may be regarded as
some approximation defined by a sequence of self-evident
assumptions that serves to simplify the homogeneous
Bethe--Salpeter equation \cite{BS}, which constitutes within the
framework of relativistic quantum field theories the adequate
approach to bound states of the fundamental degrees of freedom of
such quantum field theory:\begin{enumerate}
\item The most decisive simplification is to disregard any
reference to \emph{timelike\/} coordinate and momentum variables,
thereby defining an \emph{instantaneous\/} Bethe--Salpeter
formalism \cite{WLe}.\item Enabling any bound-state constituent to
propagate freely entails Salpeter's equation~\cite{SE}.\item
Skipping impacts of negative-energy contributions gives the
\emph{reduced\/} Salpeter equation.\item Ignoring the spins of all
bound-state constituents leads to the \emph{spinless\/} Salpeter
equation.\end{enumerate}

\section{General Insights: Elementary Constraints on Bound-State
Spectra}\label{S}Even if being not accessible analytically, the
spectrum of the operator $\widetilde H$ is tightly
constrained.\pagebreak
\begin{description}\item[1.~Boundedness from below of Hamiltonian
operators] is, as kinetic operators are positive, obvious if
$V(\bm{x})$ itself is bounded from below. If not, comparison with
the relativistic Coulomb problem may prove advantageous. The
Hamiltonian operator $\widetilde H$ with the Coulomb interaction
$$V(\bm{x})\equiv-\frac{\alpha}{|\bm{x}|}\ ,\qquad\alpha>0\ ,$$
will be bounded from below \cite{IWH1,IWH2} if and only if the
coupling strength $\alpha$ satisfies the constraint$$\alpha\le
\frac{4}{\pi}=1.273239\dots\ .$$Within this interval, the
spectrum, $\sigma(\widetilde H)$, of this operator $\widetilde H$
is subject to the lower bound~\cite{IWH1,IWH2}\begin{equation}
\sigma(\widetilde H)\ge2\,m\,\sqrt{1-\left(\frac{\pi\,\alpha}{4}
\right)^{\!2}}\ .\label{IWH}\end{equation}For Coulomb couplings
$\alpha\le1$, the lower bound (\ref{IWH}) may, straightforwardly,
be improved~\cite{MR}~to$$\sigma(\widetilde H)\ge2\,m\,
\sqrt{\frac{1+\sqrt{1-\alpha^2}}{2}}\ .$$\item[2.~Number of
discrete eigenstates in the spectrum of $\widetilde H$:]For all
potentials $V(\bm{x})$ that satisfy\begin{equation}V(\bm{x})=V(r)
\in L^{3/2}({\mathbb R}^3)\cap L^3({\mathbb R}^3)\ ,\qquad
V(\bm{x})\leq0\ ,\qquad r\equiv|\bm{x}|\ ,\label{L}\end{equation}
the number of bound states of such operator $\widetilde H$ may be
shown to be bounded from above~by~\cite{ICD}$$N\le\frac{C}
{12\,\pi}\int\limits_0^\infty{\rm d}r\,r^2\left[|V(r)|
\left(|V(r)|+4\,m\right)\right]^{3/2}\ ,$$ with the constant $C$
given by $C=14.107590867$ for $m>0$ or $C=6.074898097$ for
$m=0$~\cite{WL:K}.\item[3.~Upper bounds on the eigenvalues of
$\widetilde H$]arise as rather trivial consequence of the
so-called minimum--maximum theorem \cite{RS}: for every
self-adjoint operator $H$ bounded from below with (ordered)
eigenvalues $E_0\le E_1\le E_2\le\cdots$, the $d$ (ordered)
eigenvalues $\widehat E_0\le\widehat E_1\le\cdots\le\widehat
E_{d-1}$ of the operator $\widehat H$ defined by a restriction of
$H$ to a $d$-dimensional subspace of the domain of $H$ provide
upper bounds to the first $d$ eigenvalues of $H$ below the onset
of the \emph{essential\/}~spectrum,$$E_k\le\widehat E_k\qquad
\forall\quad k=0,1,2,\dots,d-1\ .$$Its eigenstates can be expanded
in terms of some convenient basis \cite{WL:T1,WL:T2}, e.g., in
products of generalized-Laguerre polynomials, $L_k^{(\gamma)}$,
and spherical harmonics ${\cal Y}$ for angular momentum~$\ell$:
\begin{align}\psi_k(\bm{x})\propto r^{\ell+\beta-1}\exp(-\mu\,r)
\,L_k^{(2\ell+2\beta)}(2\,\mu\,r)\,{\cal Y}(\Omega_{\bm{x}})\
,\qquad k\in{\mathbb N}_0\ ,&\label{LB}\\L_k^{(\gamma)}(x)\equiv
\sum_{t=0}^k\,\binom{k+\gamma}{k-t}\,\frac{(-x)^t}{t!}\ ,\qquad
\mu>0\ ,\qquad\beta>-\frac{1}{2}\ .&\nonumber\end{align}
\item[4.~Quality of eigenstates] \cite{WL:Q1,WL:Q2} tentatively
localized by whatever technique forms a crucial aspect that may be
quantified by the \emph{relativistic\/} generalization
\cite{WL:RVT1,WL:RVT2} of the virial theorem of nonrelativistic
quantum theory, which relates, for any of the eigenstates
$|\chi\rangle$ of the operator (\ref{H}), the expectation values
of the radial derivatives of both the kinetic terms and the
potential~$V(\bm{x})$:$$\left\langle\chi\left|
\frac{\bm{p}^2}{\sqrt{\bm{p}^2+m_1^2}}+
\frac{\bm{p}^2}{\sqrt{\bm{p}^2+m_2^2}}\right|\chi\right\rangle=
\left\langle\chi\left|\,\bm{x}\cdot\frac{\partial\,V}
{\partial\bm{x}}(\bm{x})\right|\chi\right\rangle.$$\end{description}
\pagebreak

All of this proved useful in locating the discrete spectra of
problems defined by spherically symmetric (central) potentials
$V(\bm{x})=V(r)$ (where $r\equiv|\bm{x}|$) of Woods--Saxon
\cite{WL:WS,WL:Q@W}, Hulth\'en \cite{WL:Q@W,WL:H}, Yukawa
\cite{WL:Y}, kink-like \cite{WL:K} or generalized-Hellmann
\cite{WL:Hm,WL:R} type. As an illustration, an application of all
these tools is sketched for the totality of generalized-Hellmann
potentials.

\section{Application: Class of Potentials Generalizing Hellmann's
Proposal}
\begin{figure}[ht]\centering{\begin{tabular}{cc}
\includegraphics[scale=1.7076,clip]{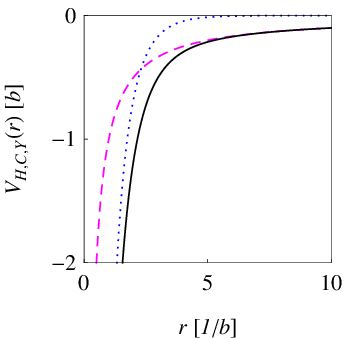}\hspace{1ex}&\hspace{1ex}
\includegraphics[scale=1.7076,clip]{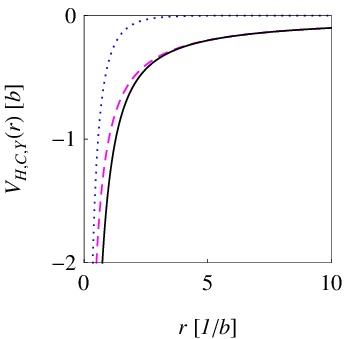}\\[2ex](a)&(b)\\[4.3849ex]
\includegraphics[scale=1.7076,clip]{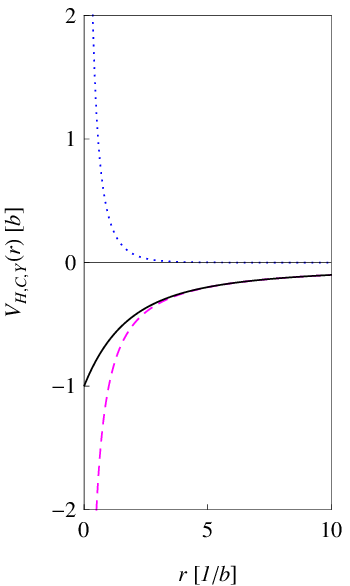}\hspace{1ex}&\hspace{1ex}
\includegraphics[scale=1.7076,clip]{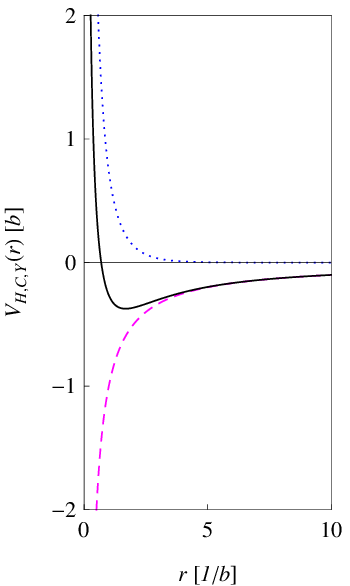}\\[2ex](c)&(d)\end{tabular}}
\\[1.52405ex]\caption{Representatives of four (out of the in total
seven) different categories of generalized Hellmann potentials
$V_{\rm H}(r)$, specified by the (in)equalities of their couplings
(a) $\upsilon>\kappa$, (b) $\upsilon=\kappa$, (c)
$\upsilon=-\kappa$, and~(d) $\upsilon<-\kappa$, obtained as sums
(solid black) of Coulombic (dashed magenta) and Yukawa (dotted
blue) terms. The systematic classification of all
\emph{generalized\/} Hellmann potentials can be found in Table~1
of Ref.~\cite{WL:R}.}\label{Pos}\end{figure}

Let us take the liberty to generalize the form proposed
by Hellmann \cite{H1,H2} to a superposition$$V_{\rm H}(r)\equiv
V_{\rm C}(r)+V_{\rm Y}(r)=-\frac{\kappa}{r}-\upsilon\,
\frac{\exp(-b\,r)}{r}$$of a definitely attractive Coulomb
potential $V_{\rm C}(r)$ and a Yukawa-inspired potential $V_{\rm
Y}(r)$, with Coulomb coupling $\kappa$, and Yukawa coupling
$\upsilon$ and slope parameter $b$, subject to the constraints
$$\kappa\ge0\ ,\qquad\upsilon\gtreqqless0\ ,\qquad b>0\ .$$Within
the obtained class of \emph{generalized\/} Hellmann potentials
$V_{\rm H}(r)$, the \emph{qualitative\/} behaviour of any of its
members depends on the relative sign and size of the two coupling
strengths $\kappa$ and $\upsilon$ that govern the Coulomb and
Yukawa contributions to such generalized Hellmann~potential. These
potentials show a significant diversity: they may be unbounded
[Fig.~\ref{Pos}(a,b)] or bounded [Fig.~\ref{Pos}(c,d)] from below,
and be singular [Fig.~\ref{Pos}(a,b,d)] or finite
[Fig.~\ref{Pos}(c)] at the origin, $r=0$.

Needless to stress, using the tools sketched in Sect.~\ref{S}, it
is a breeze to rough out the~spectra of the entirety of
relativistic Hellmann problems (for some relevant details, consult
Ref.~\cite{WL:Hm}).\begin{itemize}\item Evidently, boundedness
from below is the question to be answered first and foremost:
Since \emph{any\/} of our generalized Hellmann potentials is
trivially bounded from below by conveniently selected Coulomb
potentials, suboptimal bounds on the spectrum $\sigma(\widetilde
H)$ necessarily exist for $$\kappa+\upsilon\le\frac{4}{\pi}\
.$$For the sum of couplings non-positive, i.e., $\kappa+\upsilon
\le0$, this lower bound of Coulomb origin~is easily improvable:
$V_{\rm H}(r)$ is bounded from below and hence $\sigma(\widetilde
H)$ by the minimum of $V_{\rm H}(r)$.\end{itemize}

\begin{table}[hb]\centering\caption{Binding energies $E_k-2\,m\;[m]$
($k\in{\mathbb N}_0$) of the spinless relativistic Hellmann
problem \cite{WL:Hm,WL:R}, for lowest-lying states of radial
excitation $n_{\rm r}$ and orbital angular momentum $\ell$,
assuming $b=\mu=m$ and $\beta=1$ for the potential-slope and
Laguerre-basis parameters, and trial-space dimensions around
$d\gtrapprox29$.}\label{B}\begin{tabular}{cclll}\toprule&&
\multicolumn{3}{c}{Upper bounds}\\\cline{3-5}\\[-2ex]
\multicolumn{2}{c}{Bound state}&\multicolumn{1}{c}
{$\kappa=\upsilon=\frac{1}{2}$}&\multicolumn{1}{c}{$\kappa=1$,
$\upsilon=-1$}&\multicolumn{1}{c}{$\kappa=1$, $\upsilon=-2$}\\
\cline{1-2}\\[-2ex]$n_{\rm r}$&$\ell$&\multicolumn{1}{c}{[case
Fig.~\ref{Pos}(b)]}&\multicolumn{1}{c}{[case Fig.~\ref{Pos}(c)]}&
\multicolumn{1}{c}{[case Fig.~\ref{Pos}(d)]}\\\midrule
0&0&$\quad\!\!-0.11673$&$\quad-0.17951$&$\quad-0.14410$\\
0&1&$\quad\!\!-0.01579$&$\quad-0.06294$&$\quad-0.06157$\\
0&2&$\quad\!\!-0.00616$&$\quad-0.02813$&$\quad-0.02812$\\[.8ex]
1&0&$\quad\!\!-0.02107$&$\quad-0.05464$&$\quad-0.04786$\\
1&1&$\quad\!\!-0.00509$&$\quad-0.02810$&$\quad-0.02762$\\[.8ex]
2&0&$\quad\!\!-0.00688$&$\quad-0.02566$&$\quad-0.02338$\\
\midrule\multicolumn{2}{l}{Lower bound}&$\quad\!\!-0.58578\dots$&
$\quad-1$&$\quad-0.37336\dots$\\\bottomrule\end{tabular}\end{table}

\clearpage

\begin{itemize}\item The number of discrete eigenstates of
\emph{any\/} generalized-Hellmann Hamiltonian may never~be bounded
from above because any generalized Hellmann potential becomes
Coulomb-like at large distances, in the limit $r\to\infty$. More
technically, any generalized relativistic Hellmann problem fails
\cite{WL:Hm} to satisfy the prerequisite (\ref{L}) that guarantees
finiteness \cite{ICD} of the number.\item Upper bounds on the
eigenvalues corresponding to lowest-lying discrete eigenstates may
be conveniently calculated numerically upon spanning that
$d$-dimensional subspace referred to by the minimum--maximum
theorem by a Laguerre basis (\ref{LB}), as exemplified in
Table~\ref{B};~they can be optimized by increasing the dimension
$d$ and adapting the two parameters $\mu$
and/or~$\beta$.\end{itemize}

\end{document}